\documentclass{aa_preprint}

\usepackage{color}
\usepackage[varg]{txfonts}
\usepackage{amsmath}

\usepackage{graphicx}

\usepackage[breaklinks=true]{hyperref}

\begin{document}

\title{Rotational radial shear in the low solar photosphere
 }
 \subtitle{Direct detection from high-resolution spectro-imaging}

  \author{
  T. Corbard\inst{1}
   \and M. Faurobert\inst{1}
   \and   B. Gelly\inst{2}
   \and R. Douet\inst{2}
   \and D. Laforgue\thanks{Deceased}\inst{2}
  }

   \institute{Universit\'{e} C\^{o}te d'Azur,  Observatoire de la C\^{o}te d'Azur, CNRS, Laboratoire Lagrange, Bd de l'observatoire, CS34229, 06304 Nice cedex4, France\\
              \email{thierry.corbard@oca.eu, marianne.faurobert@oca.eu}     
              \and      CNRS-IRL2009, c/o IAC Via Lactea s/n,  ES38205 La Laguna, Tenerife, Spain \\
              \email{bgelly@themis.iac.es}
              }

\date{Accepted 22 August 2025}
\titlerunning{ Rotational shear in the low solar photosphere  }
 \authorrunning{Corbard et al.}

\abstract
{
Radial differential rotation is an important factor in stellar dynamo theory. In the Sun, helioseismology has revealed a near-surface shear layer in the upper 5–10 percent of the convection zone. At low to midlatitudes, the rotation velocity gradient decreases sharply near the surface. A depth gradient in rotational velocity was recently detected in the low photosphere using a differential interferometric method on spectroscopic data. Granular structures at different depths in the Fe~I 630.15~nm line showed a systematic retrograde shift compared to continuum structures, which suggests a height-related decrease in angular velocity. This estimate depends on the assumed granulation coherence time.}
{ We use a more direct approach to measure the differential rotational velocity at different photospheric heights.}
{We performed spectroscopic scans of the same granular region in Fe~I 630.15~nm and Ca~I 616.2~nm lines, and measured displacements of images at different line chords between consecutive scans. These observations require excellent seeing, stable adaptive optics correction, and scanning times shorter than the granulation lifetime.  Adaptive optics stabilizes continuum images but not higher-altitude rotation differences.   We used both THEMIS and HINODE Solar Optical Telescope Fe~I 630.15~nm data to measure formation height differences via perspective shifts observed away from the disk center with the slit radially oriented.}
{ Measurements at disk center and $\pm 25 ^{\circ}$ latitude along the central meridian show a parabolic decrease in rotational velocity with height that reaches about 16\% slower rotation at 80~km above the continuum. No significant difference is found between equator and $\pm 25 ^{\circ}$ latitudes.}
{ The low photosphere is a transition zone between the convective and radiative layers. Our measurements provide new constraints on its dynamical behavior and valuable boundary conditions for numerical simulations of the Sun’s upper convection zone.}


\maketitle

\section{Introduction} 
\label{Sec:Intro}
Rotational shear plays a major role in the onset of stellar dynamo. In the Sun, helioseismic observations revealed two regions that show a strong height gradient of the rotation rate: the tachocline at the base of the convective zone and the near-surface shear layer (NSSL).
More recent global and local helioseismology results \citep[e.g.,][]{zaatri2009,reiter2020,komm2022} covering longer periods and a full solar cycle confirm that the gradient of rotation is not uniform in the NSSL,  but that the rotation rate decreases more steeply close to the surface
\citep[see][Fig.~5]{komm2022}. 

This paper aims to provide new observational constraints on the rotational profile in the transition layer between the NSSL of the convective envelope and the radiative solar photosphere. 
In a recent paper devoted to radiative hydrodynamic numerical simulations of the solar near-surface in the presence of rotation, \citet{kosovichev2023} derive a strong negative gradient of rotation, $\partial \ln\Omega/\partial \ln r \simeq -4,$ in a subsurface layer at depths smaller than 4~Mm.  
 However, the numerical noise increases drastically near the surface. Measurement of the rotational profile in the low photosphere would provide reliable boundary conditions for this kind of simulation.
Moreover, the interface formed by the NSSL and the low photosphere is a boundary layer for the solar convection problem. Hydrodynamical simulations of the solar convective envelope are very challenging. The convective velocities that are required to transport the solar luminosity in the most recent global models of convection are systematically too large to explain the differential rotation measured by helioseismology \citep{hotta2023}.  We believe that observational constraints on the rotational profile at the interface between the NSSL and the photosphere would help improve global simulations of solar convection.

Current magnetohydrodynamical simulations of the photosphere do not account for any depth-dependent gradient in the rotation rate. Since the turnover time of the granulation pattern of about 10 minutes is much shorter than the solar rotation period, the convective motions in the photosphere are practically unaffected by the Coriolis force. However, the impact of steep rotational shear on the local dynamo, which generates ubiquitous small-scale magnetic fields, is not clear. 

Using a new analysis of 5-mn solar p-mode limb oscillations from 3.5 years of Helioseismic and Magnetic Imager (HMI) data, \citet{kuhn2017} measured a steep radial decrease of the latitudinal average of the solar rotation rate in the upper photosphere. They argue that angular momentum loss could be driven by the torque due to the escape of photons from the photosphere.
 In a recent work \citep[Paper I in the following]{ corbard2024, faurobert2023}, a steep rotational velocity gradient was detected in the low photosphere. The detection is not based on the Doppler effect but on measuring systematic retrograde shifts between images of the coherent granular pattern taken simultaneously at different altitudes in the wings of the Fe~I 630.15~nm line.  The formation height of the images at different line chords was derived from the perspective effect using scans performed off the disk center with the spectrograph slit oriented radially.  The observations were done at the solar telescope THEMIS equipped with an efficient adaptive optics (AO) system. Retrograde shifts were interpreted as a consequence of the horizontal dragging of the structures by the rotational gradient. They depend on how long the velocity gradient has acted to drag the upper layers, i.e., on the coherence time of the granulation pattern.  Thus, estimating the rotational depth gradient from the measured shifts is highly sensitive to the coherence time of the signal-producing structures.

Here, we propose to use a different approach. By performing a series of spectroscopic scans of a granulation scene, we can measure the shift along the slit of the granulation images seen at different line chords between consecutive scans. The measurement would be performed on all the slit positions of the scan and on several scans, yielding the bulk velocity in the field of view along the direction of the slit. The AO system corrects for the image motion at the continuum level. However, if the velocity varies with height,  the line-wing images formed higher up in the photosphere would still appear shifted between two consecutive scans. As the scanning time is known, we can obtain the difference of the velocity between the two layers without any assumption on the coherence time of the granulation. We still need to scan the field of view fast enough as compared to the lifetime of the granules. In the following, we test these ideas on spectroscopic scans performed with the spectrograph slit both along the west-east direction and along the south-north direction, which allows us to explore the differential azimuthal and meridional velocities, respectively.
To measure the shifts, we compute the phase of the Fourier cross-spectrum of the intensity distributions for a given line chord along the spectrograph slit in two consecutive scans, and derive its slope. 
We emphasize that this method requires very good and stable seeing conditions, as consistent image quality must be maintained while tracking the same granular scene across several spectroscopic scans.

A similar method was previously implemented by \citet{habbal2021} using space-borne observations of granulation in the 617.3~nm  continuum acquired with HMI aboard the Solar Dynamics Observatory \citep{Scherrer2011}. In this work, the latitudinal differential rotation was obtained by measuring the local shift, at the central meridian, between two images separated by a given time interval. The measurement method relies on an iterative phase correlation technique with sub-pixel accuracy.
As in Paper I, we derive the formation height of the line-wing images from the measurement of their perspective shift with respect to the continuum image, when the granular scene is observed away from the center of the solar disk with the spectrograph slit oriented radially.

In the following section, we present our observations and methods. Section 3 is devoted to the measurement of the perspective shift from THEMIS and HINODE spectroscopic data in the Fe~I line at 630.15~nm and from THEMIS data in the Ca~I 616.2~nm line  (not observed aboard HINODE). In Section 4 we implement the method outlined above on some very good observing sequences recorded at THEMIS in July 2023; we derive the differential azimuthal velocity in the low photosphere up to the altitude $z= 80$~km. The meridional velocity gradient is also investigated using a sequence recorded at the center of the solar disk with the slit oriented south-north; no significant differential velocity is detected in this sequence over this altitude range. In the last section, we draw some conclusions and perspectives.

\section{Observations and methods}
\label{sect2}

We obtained spectroscopic observations at the THEMIS solar telescope, from July 25 to 29, 2023, in the Fe~I 630.15~nm and Ca~I 616.2~nm lines, processed following \citet{wohl2002}. Complementary HINODE Solar Optical Telescope spectropolarimeter (SOT-SP, \cite{Tsuneta2008}) data from the irradiance program HOP79 (July 22, 2023) were also analyzed.

At THEMIS, 40\arcsec\ × 108\arcsec\ quiet-Sun regions were scanned at several locations along the solar polar axis, with the spectrograph slit alternately oriented east–west and north–south. The slit width was 0.5\arcsec\  and the step size 1\arcsec. HOP79 performs monthly scans of 30\arcsec\ × 130\arcsec\ regions at 20 positions along the polar axis, spanning from the north to the south limb, with the slit aligned north–south.

Spatial sampling along the slit was 0.235\arcsec\ for THEMIS and 0.32\arcsec\ for SOT-SP; spectral pixels were 1.35~pm and 2.1~pm, respectively. HINODE avoids seeing effects entirely, while at THEMIS we benefited from excellent conditions and adaptive optics \citep{thiebaut2022,gelly2016}. Over the five observing days, the best seeing occurred between 07:00 and 11:00 UT. All scans analyzed were taken within this interval and only those with high continuum granulation RMS contrast were retained. 

The displacement measurement technique, detailed in Paper~I, uses the phase of the cross-spectrum between two spectroscopic images, computed in Fourier space, to detect subpixel shifts. Because the displacement was aligned with the slit, only one-dimensional intensity profiles were analyzed. Power spectra of the intensity variations were calculated at each slit position and averaged over the entire scan to improve the signal-to-noise ratio. This process was repeated for multiple line chords to sample different formation heights in the photosphere.
This method is based on the assumption that the images are similar but exhibit slight shifts. It is not constrained by the instrument’s spatial resolution, and the primary limitation lies in the signal-to-noise ratio of the image spectrum.

When we apply this technique to granulation images that contain a large number of granular structures taken at the same time in the wing of a spectral line and in the continuum with a radially oriented slit, the displacement, $\delta,$ is due to the perspective effect because the images are formed at
different altitudes. Unlike stochastic motions, this systematic effect is not eliminated by averaging over numerous realizations of the granulation pattern. The formation height, $z,$ of the image above the continuum level is related to $\delta$ by $\delta=z\sin(\theta)$ where $\theta$ is the heliocentric angle; its sign changes for symmetrical positions with respect to the disk center.

Here, we also propose to apply this method to granulation images taken at the same level in a spectral line but at different times, namely between two successive scans, to measure their shift along the spectrograph slit during the duration of a scan. The scans are performed on targets located along the central meridian of the Sun with the spectrograph slit along the west-east and along the south-north directions. When the slit is along the west-east direction, the value of $\delta$ directly gives the image displacement along this direction without any projection effect.
When the spectrograph slit is in the south-north direction, the displacement $dy$ during the duration of a scan is related to $\delta$ by $\delta =dy\cos\theta$.

As mentioned above, the AO system corrects for image motion in the continuum during scans. However, the correction is not perfect; there is always some residual motion of the field of view between two consecutive scans. So, in this paper, we do not pretend to measure the rotational or meridional velocity;  our aim is to explore systematic differential motions along the spectrograph slit of images formed at different altitudes in the photosphere with respect to the image taken in the continuum. 
 In the following, we  test this idea on granulation images taken at different line chords. The measurement of the granulation motions along the slit relies on the phase of the cross-spectrum between the intensity distribution at a given line chord at two successive scans. Comparing similar images requires the field of view to remain sufficiently stable enough between the two scans, i.e., the displacement perpendicular to the slit from AO correction residuals must be smaller than the width of the slit. We checked that this condition is fulfilled for the observing sequences that we have selected to carry out this experiment.

To implement these techniques, our first step is to generate maps that correspond to surfaces of approximately uniform continuum optical depth. The reconstruction procedure exploits the connection between line and continuum absorption in the Lorentzian wings of spectral lines, as described in detail by \citet{faurobert2012, faurobert2023}. In this work, we apply the approach to observations from both SOT-SP and THEMIS, focusing on the Fe~I 630.15~nm and Ca~I 616.2~nm lines. The resulting data consist of 25 line chords, distributed evenly in width from the line core (chord 1) to the far wing at 0.9 $I_c$ (chord 25), the latter serving as the reference map in the subsequent analysis.
The analysis presented below was performed on scans of 40\arcsec\ in the scan direction and 47\arcsec\ (200 pixels) along the slit, centered on the middle of the scanned region.

\section{Perspective shift measurement from HINODE and THEMIS data}
\label{Sect2}

We used spectroscopic scans taken along the polar axis with the spectrograph slit oriented in the south-north direction.  Before computing the cross-spectra between line-chord and continuum images, we applied a Gaussian filter with a width of 0.8\arcsec\  to the spectrograms along the slit to reduce phase noise that appears at high spatial frequencies. We verified that this filtering does not significantly affect the phase-slope measurement, which is performed over a reduced spatial frequency domain, i.e., on large-scale structures (see below). 

   Figure~\ref{fig1} shows the phase of the cross-spectrum of the continuum image with the image at line chord 15 in the Fe~I 630.15~nm line, with both SOT-SP (green symbols) and THEMIS (blue and red symbols) for $\cos\theta=0.81$ in the northern and southern hemispheres.
 At spatial frequencies below 0.6~arcsec$^{-1}$, the phase exhibits a roughly linear trend, corresponding to structures larger than 1.7\arcsec. A linear fit over this range, shown in the figure, reveals the expected perspective effect: the phase slope is negative in the northern hemisphere and positive in the southern hemisphere. Data from HINODE and THEMIS show good agreement. Figure~\ref{fig2}  shows the same quantities for the Ca~I 616.2~nm line; however, this line is not observed on HINODE.
 
 In Paper I, we investigated the depth variation of the rotational velocity at heights up to $z=32$~km only. The measurement of the perspective shift requires the images at the two levels to be sufficiently similar. In the present work, we reached higher altitudes by proceeding in two steps. We first measured the perspective effect between the continuum level and the line-chord images up to line chord 15. Then we measured the perspective shift at smaller line chords with respect to the image at line chord 15 and we derived their formation height by summing the two shifts. The method is still limited to altitudes below the granulation contrast inversion layer, where the image contrast becomes close to zero; the cross-spectrum phase then becomes too noisy for a reliable measurement. 
 
 Figure~\ref{fig3} shows the formation height of the images at the different line levels obtained from the measurement of the perspective shifts in the northern and southern hemispheres for the Fe~I 630.15~nm line up to line chord 10 and for the Ca~I 616.2~nm line up to line chord 6. The standard deviation on the measurement is derived from the standard deviation of the estimate of the slope of the linear fit. It increases at line chords smaller than 15 because, in the two-step procedure, we also have to sum the variances of the two measurements. 
 Tables \ref{tab1} and \ref{tab2} present the formation heights, $z,$ of the images at different line chords derived from the average of the measurements in both hemispheres for the Fe~I line and the Ca~I line, respectively.   
 The formation heights given in Tables 1 and 2 are measured with respect to the formation level of the nearby continua, at 630~nm and 616~nm, respectively. We computed the optical depth $\tau_{630}$ and $\tau_{616}$  of the two continua in a standard one-dimensional model of the solar photosphere (model 1001 of \citet{fontenla2011}).  Both continua have an optical depth equal to unity at $ z\simeq 7$~km above the formation level of the 500~nm continuum. Therefore, the height difference between the two continuum formation levels is not significant compared to the accuracy of the measurements given in the tables. In the following, we assume that both continua are formed at the same height.

\begin{figure}[ht]
\includegraphics[width=0.9\columnwidth]{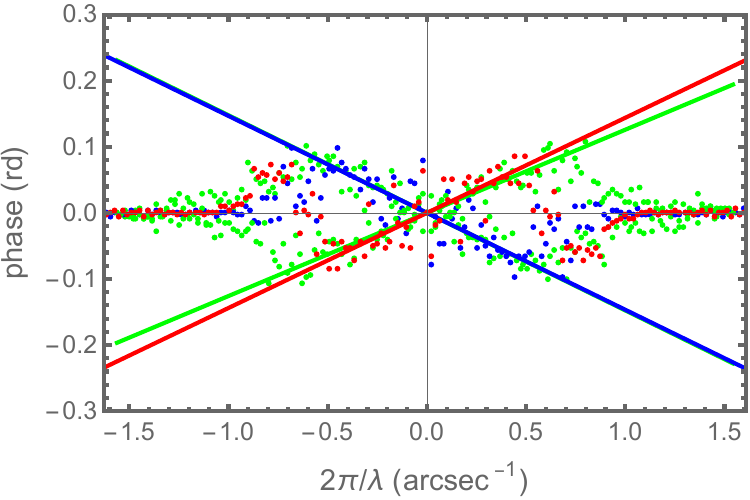}
 \caption{Phase of the cross-spectrum of the images at the reference level and at line chord 15 in the Fe~I 630.15~nm line as function of the spatial frequency.
 The spectrograph slit is oriented along the north-south axis at $\cos\theta= 0.81$. Red (blue) symbols: THEMIS data in the southern (northern) hemisphere. Green symbols: HINODE data at the same heliocentric angle in both hemispheres.
 }
  \label{fig1}  
\end{figure} 
\begin{figure}[ht]
\includegraphics[width=0.9\columnwidth]{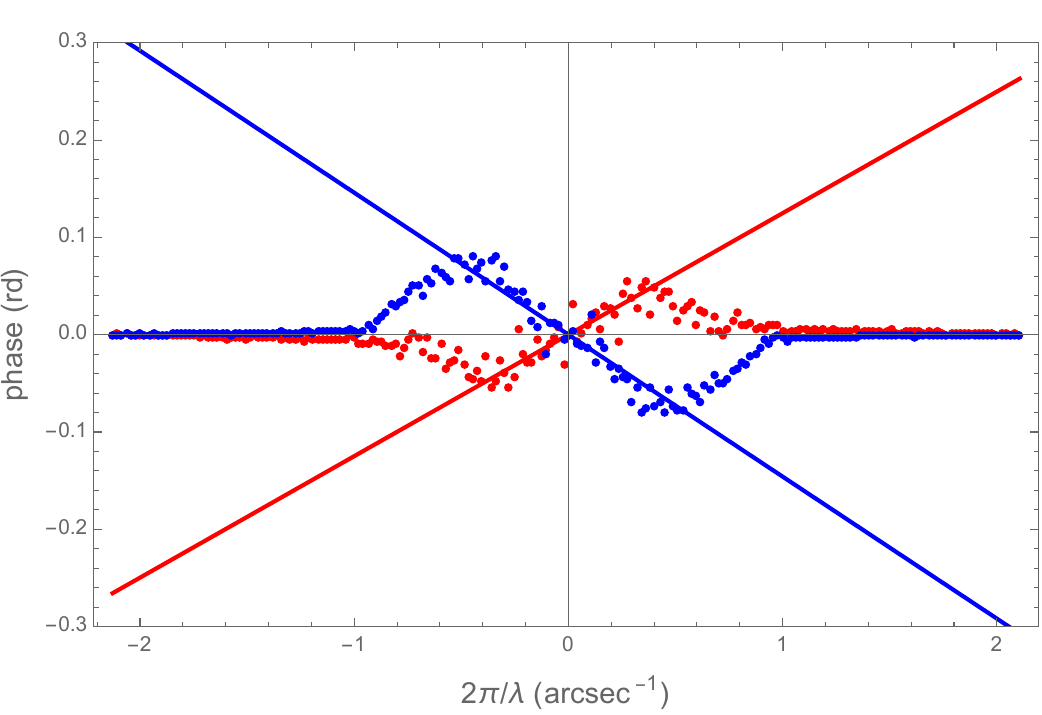}
 \caption{Same as Fig. \ref{fig1} but for level 15 of the Ca~I 616.2~nm line. No HINODE data is available for this line.
 }
  \label{fig2}
\end{figure} 
\begin{figure}[ht]
\includegraphics[width=0.9\columnwidth]{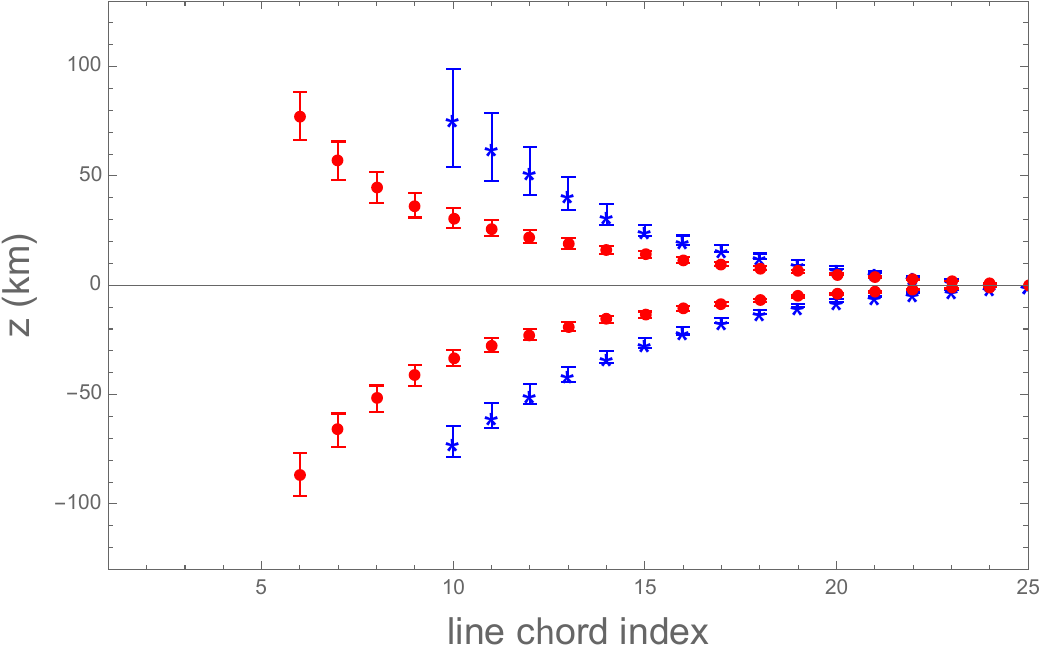}
 \caption{Formation heights (in kilometers) of the images at the different line chords derived from the perspective effect in both hemispheres. Positive values are obtained in the northern hemisphere where the perspective displacement is toward the north, and negative values in the southern hemisphere. 
Blue symbols:  Fe~I 630.15~nm line. Red symbols: Ca~I 616.2~nm line. }
  \label{fig3}
\end{figure} 

\begin{table*}
\caption{Average value of the formation height of the line-chord images for the Fe~I 630.15~nm line above the 630~nm continuum level.}            
\label{tab1}     
\centering                                      
\begin{tabular}{c c c c c c c}          
\hline\hline                       
line chord &$20$& $19$ &$ 18$&$17$&$16$&$15$\\ 
\hline                                   
z (km) & $7.6 \pm 0.6$ & $9.9  \pm 0.9 $ & $12.8 \pm 1.1$ & $16.4\pm 1.5$& $20.8\pm 1.9$& $25.7\pm 2.4$\\
\hline
line chord &14&13&12&11&10\\ 
\hline
 z (km) & $32.6\pm 3.9$ & $41.3 \pm 5.9$ & $51.0 \pm 8.4$& $61.4\pm 11.8$& $74.0\pm16.5$\\   
 \hline                              
\end{tabular}
\end{table*}

\begin{table*}
\caption{Average value of the formation height of the line-chord images for the Ca~I 616.2~nm line above the 616 nm continuum level.}             
\label{tab2}     
\centering                                      
\begin{tabular}{c c c c c c c}          
\hline\hline                        
line chord &$20$& $19$ &$ 18$&$17$&$16$&$15$\\ 
\hline                                  
z (km) & $4.6 \pm 0.5$ & $5.8  \pm 0.6 $ & $7.4 \pm 0.8 $ & $9.2\pm 1.0$& $11.1\pm 1.2$& $13.7\pm 1.5$\\
\hline
line chord &14&13&12&11&10&9\\
\hline
 z (km) & $16.0\pm1.8$ & $18.9 \pm2.3$ & $22.4 \pm 2.8$&$26.6\pm3.4$ &$32.0\pm4.2$& $38.9\pm5.2$.\\   
 \hline
 line chord&8&7&6\\ 
 z (km)& $48.3\pm6.5$&$61.7\pm8.2$&$82.0\pm 10.5$\\
 \hline

\end{tabular}
\end{table*}

\section{Measurement of the height gradient of the azimuthal and meridional velocities in the quiet photosphere}

In this section, we implement the method outlined in the Introduction to see whether it allows us to measure the gradients of large-scale velocities in the photosphere.  We restrict our study to the low photosphere, where we have been able to measure the formation heights of the line-chord images, i.e., at altitudes $z < 80$~km. 
We selected observation sequences performed under very good and stable seeing conditions. 
The selection criteria are based on the value and stability of the continuum intensity contrast along the spectrograph slit during the whole sequence.  We then checked that the displacement of the field of view in the direction perpendicular to the slit between two consecutive scans remains smaller than the width of the slit. In all the cases studied below, it remains smaller than 0.12\arcsec\  (we measured it on continuum images using the cross-spectrum phase of the intensity distribution in the direction perpendicular to the slit at two consecutive scans). As the width of the slit is 0.5\arcsec, the same granular structures appear along the slit in two consecutive scans, which allows us to measure their displacement along the slit. We then averaged this displacement over the series of successive scans. 

\subsection{Height gradient of the azimuthal velocity at the center of the solar disk}

\begin{figure}[ht]

\includegraphics[width=0.9\columnwidth]{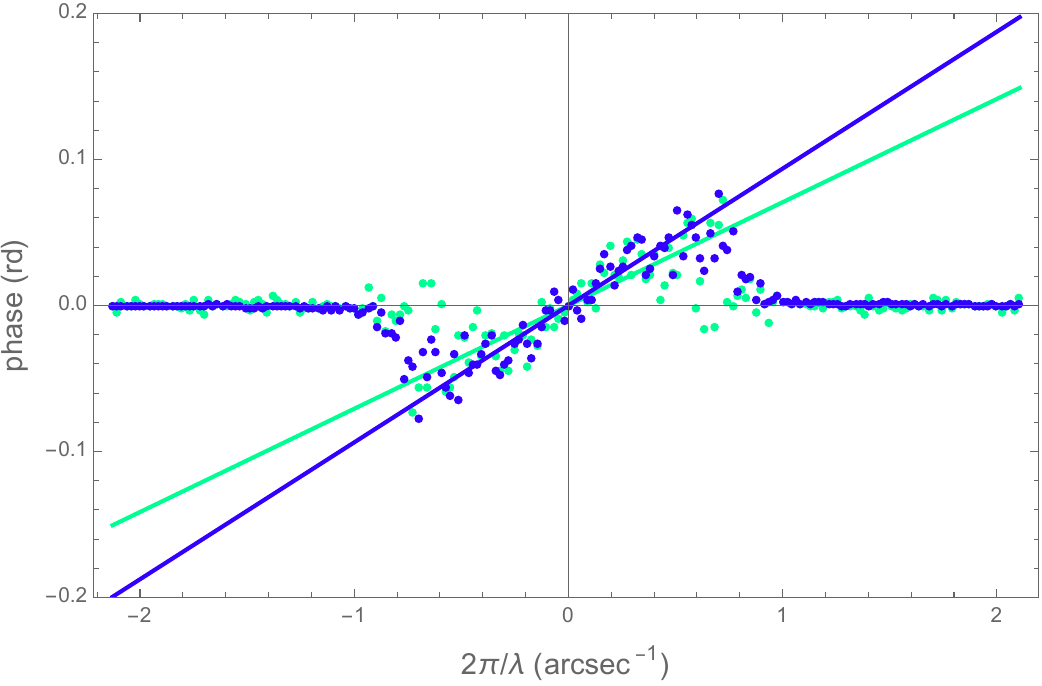}

\caption{Phase of the cross-spectra of images at two successive scans at line-chord level 12  (green symbols) and at the continuum level (blue symbols). 
}
  \label{fig5}
\end{figure}

\begin{figure}[ht]
\includegraphics[width=0.9\columnwidth]{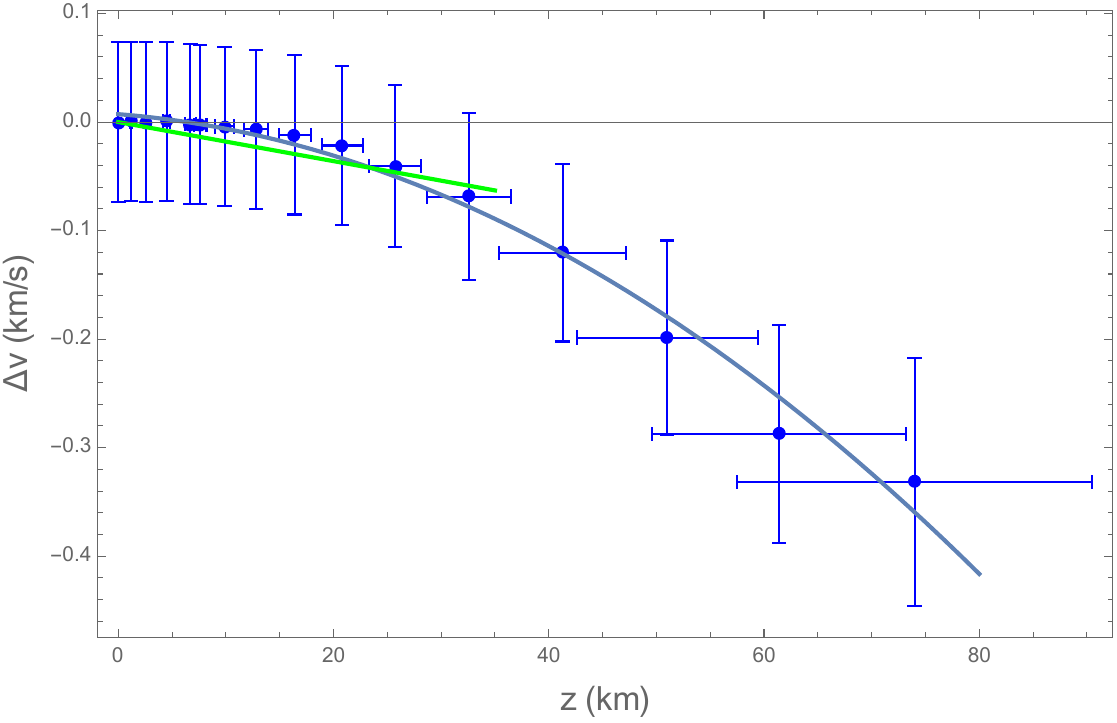}
 \caption{Differential rotation velocity measured at the center of the solar disk in the Fe~I 630.15~nm line as a function of the formation height of the line-chord images up to line chord 10. 
 The green line shows the linear fit obtained in Paper I for $0 < z < 32$~km, the blue line is a parabolic fit for $0 < z < 80$~km.}
  \label{fig7}
\end{figure}

A very good sequence of 16 successive scans was performed at the center of the solar disk with the spectrograph slit oriented along the solar equator on July 26, 2023 at 07:50~UT.
Each scan has 40 steps of 1\arcsec\  and a duration of 13.6~s.
Figure \ref{fig5} shows the phase of the cross-spectrum of the images at two consecutive scans at line chords 12 (line wing)  in the Fe~I 630.15~nm line and at the continuum level. The shifts of the images between two scans are derived from the slope of the linear part of the phase. We observe that the slope is
positive, which means that the shift is in the negative direction. The equatorial axis was oriented from the west to the east, so the shifts are westward. The rotation of the continuum image is not perfectly compensated for by the AO system, but we are primarily interested in the differential motion of the layers above the continuum level. 
By subtracting the shift of the continuum image and dividing by the duration of a scan, we obtain the differential velocities shown in Fig.~\ref{fig7} as a function of the formation heights of the line-chord images obtained in Sect.~\ref{Sect2} up to line-chord 10. Negative values mean that the layers located above the continuum formation layer rotate at a slower rate than the granulation at the continuum layer.  We observe a clear decrease in rotational velocity with height, despite rather large error bars on both the formation heights and on the relative velocities at the different heights. The error bars on $z$ and $\Delta v$ are derived from the standard deviation of the cross-spectrum phase estimate used to measure the perspective effect and the image shifts between two scans, respectively.

 In Paper I, we measured the drag of the upper layers with respect to the continuum level up to $z = 30$~km. The method we used was quite different; the cross-spectra were performed on a selection of images taken rigorously at the same time in the continuum and in the line wings to obtain a statistical average of the drag of the upper layers. Assuming a value for the coherence time of the granular pattern, we could derive a value for the rotational velocity gradient. Here, however, we independently measure the shifts of the line-wing images and of the continuum images between two scans and then
derive the differential shift caused by differential rotation during the duration of the scan. We note that the precision of the differential shift measurements below $z= 30$~km is quite poor because it involves subtracting two measurements of very similar small quantities. We would need to perform more than 16 successive scans to increase the accuracy of the measurements.  
 
 In Paper I, we observed a linear decrease in rotational velocity at altitudes below $z= 30$~km.  We expressed the rotational velocity as $\Omega(R+z)=\Omega(R)(1-\alpha z)$, where $R$ is the solar radius at the height of continuum formation. At altitude $z$, this gives $\Omega(R+z)-\Omega(R)= -\Omega(R)\alpha z$, that is, $dv(z)= -V_{rot}(R)\alpha z$. The value $\alpha \simeq 9. \, 10^{-4}$ km$^{-1}$ was obtained under the assumption that the coherence time of the granulation is $\tau =440$~s and with $V_{rot}(R)= 2$~km~s$^{-1}$.   In Fig. \ref{fig7} we show the linear decrease of Paper~I to compare it with the present results.
 We notice that both results are consistent for $z < 30$~km, but that at higher altitudes the measurements are better fitted by a parabolic fit.  The height gradient of the rotational velocity increases significantly above $z=30$~km up to $z=80$~km.

\begin{figure}[ht]
\includegraphics[width=0.9\columnwidth]{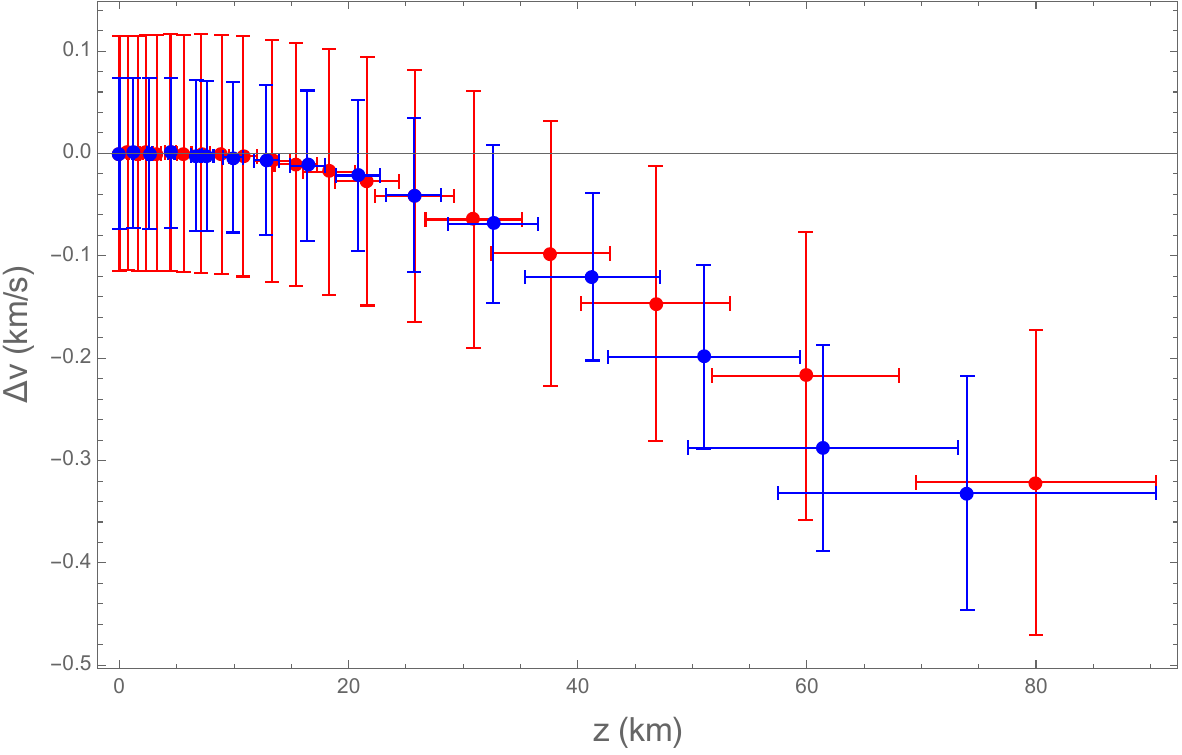}
 \caption{Comparison of the differential rotational velocities obtained at disk center with both lines. Blue symbols: Fe~I 630.15~nm line. Red symbols: Ca~I 616.2~nm line. }
  \label{fig9}
\end{figure}

In Fig.~\ref{fig9} we present  the differential velocity obtained with the same method applied to the Ca~I 616.2~nm line that was observed simultaneously on July 26, 2023 at THEMIS, and we compare it with
the results obtained with the Fe~I 630.15~nm line. We notice a fairly good agreement, but the differential velocity measurements made in the Ca~I line have larger error bars.

\subsection{Height gradient of the azimuthal velocity at latitudes  $\pm 25^{\circ}$ }

We also selected two very good sequences recorded the same day at 07:56 UT and 08:02 UT on targets located at latitudes $\pm 25^\circ$ along the polar axis with the slit parallel to the solar equator. The duration of one scan was 16.08~s and 16.4~s, respectively. 

\begin{figure}[ht]
\includegraphics[width=0.9\columnwidth]{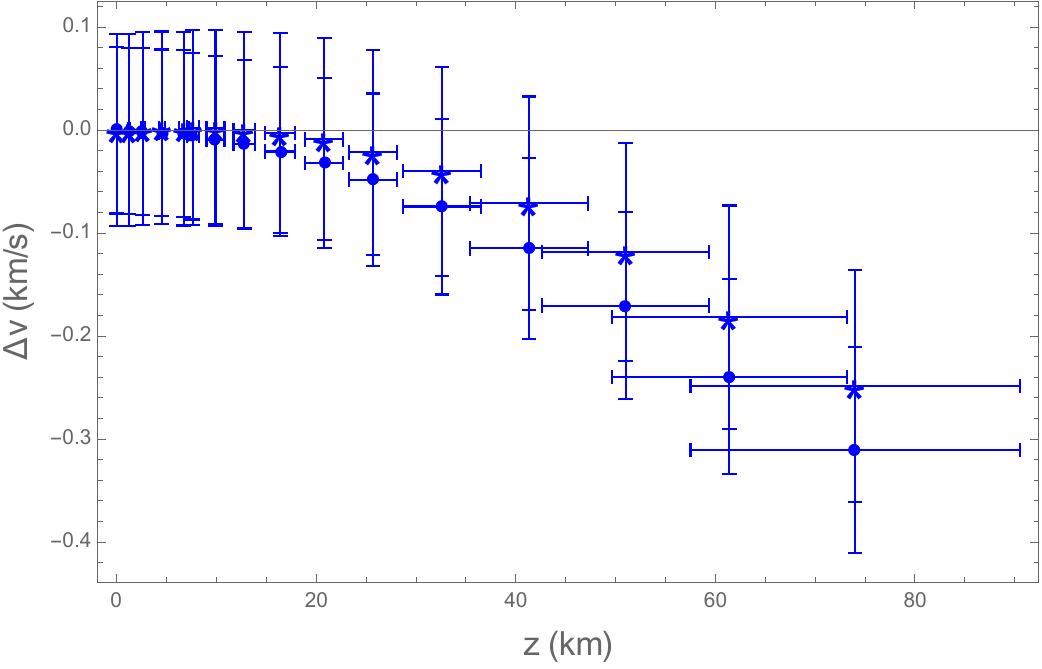}

\caption{Comparison of the differential rotational velocity obtained at $+25^{\circ}$ (stars) and $-25^{\circ}$ (dots) with the Fe~I 630.15~nm line on July 26, 2023.
}
 \label{fig15}
\end{figure}

The same analysis as for the target at disk center allows us to measure the differential shift between two consecutive scans at the different line chords and to derive the differential velocities. The results obtained at the two symmetrical latitudes $+ 25^\circ$ and $-25^\circ$ in the  Fe~I 630.15~nm line are shown in Fig. \ref{fig15}.
The behavior of the height variations of the azimuthal velocity seems quite similar to what we obtain at the center of the solar disk, but the accuracy of our measurements may not be sufficient to detect any differences.

\begin{figure}[ht]
\includegraphics[width=0.9\columnwidth]{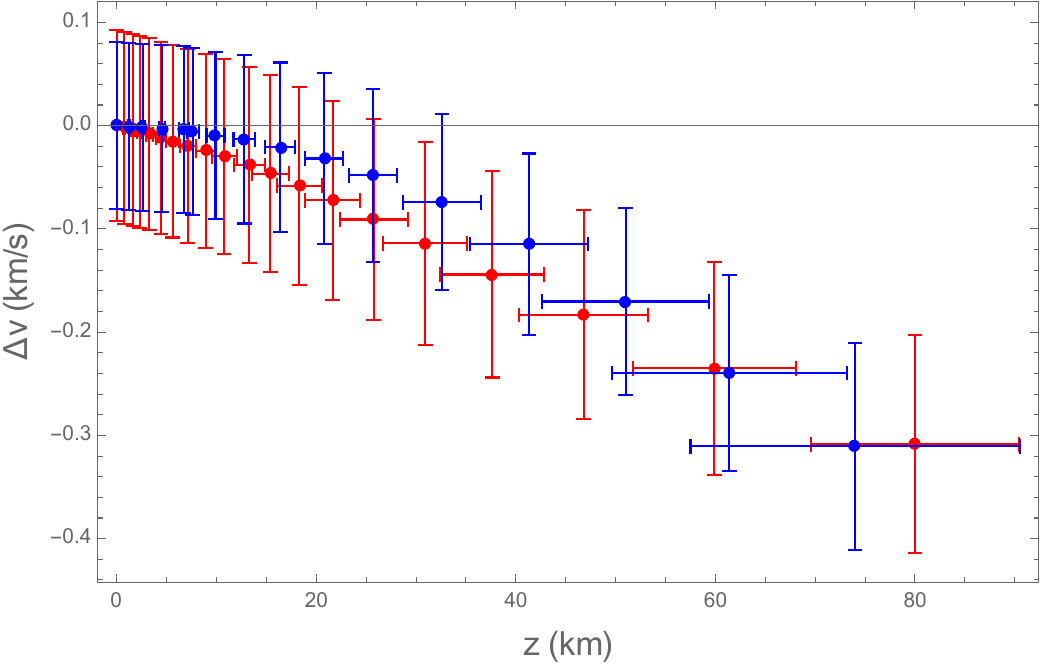}

\caption {Comparison of the differential rotational velocities obtained with both lines at -25 $^\circ$. Blue symbols: Fe~I 630.15~nm line. Red symbols: Ca~I 616.2~nm line.
}
  \label{fig13}
\end{figure}

The same analysis is performed with the Ca~I 616.2~nm line.  In Fig. \ref{fig13} we compare the differential velocity as a function of the altitude obtained with
the two lines for the target at $-25^\circ$; the results for both lines are similar within the error bars. 

\subsection{Height gradient of the meridional velocity}

\begin{figure}[ht]
\includegraphics[width=0.9\columnwidth]{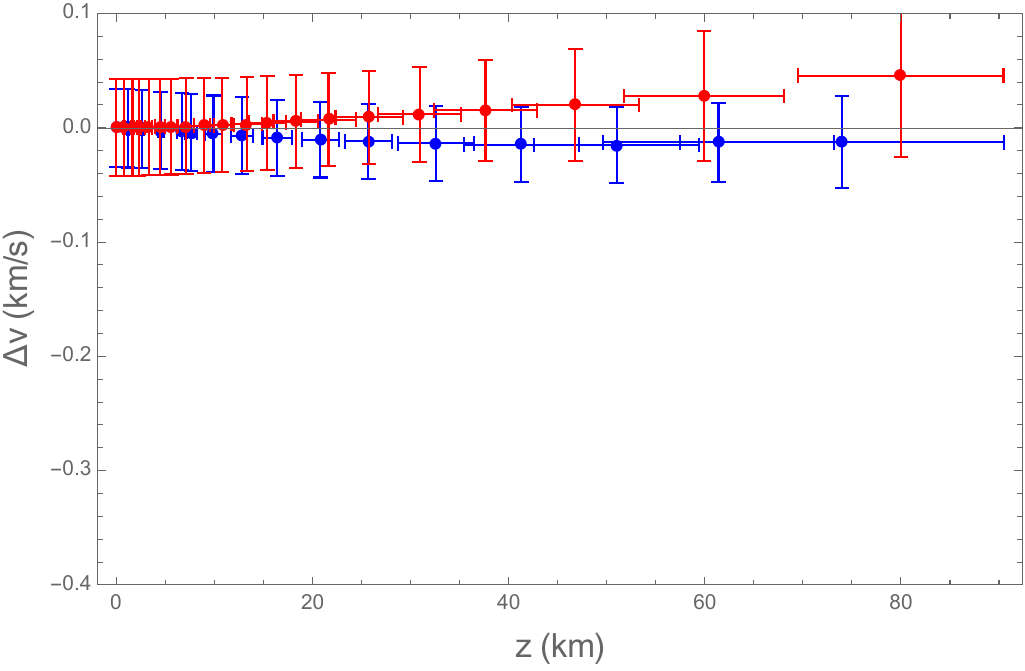}

\caption {Comparison of the differential meridional velocity  obtained with both lines at the center of the solar disk. Blue symbols: Fe~I 630.15~nm line. Red symbols: Ca~I 616.2~nm line.
}
 \label{fig16}
\end{figure}

In Paper I, we measured no drag of the granular structures at the center of the solar disk along the south-north direction. The meridional velocity is two orders of magnitude smaller than the rotational velocity in the photosphere \citep{roudier2012}. The accuracy of the measurements we present here is clearly not good enough to measure such low values. 
It is important to verify that we do not measure any significant velocity gradient along the south-north direction. This serves as a test for instrumental effects that could produce spurious velocity gradient detections in our procedure.
We use a sequence of 11 scans performed at the center of the solar disk on July 28, under very good and stable seeing conditions.  The spectrograph slit was along the south-north polar axis. As shown in Fig.~\ref{fig16}, we measure no significant differential velocity along this axis in the low photosphere up to $z= 80$ km. 
We measure no effect when none is expected, which can be considered a test of the zero level in our experiment.

\section{Conclusion}

In this work, we have tested a new method to investigate the height variations of large-scale velocities in the low photosphere of the Sun. Based on a statistical analysis of high-resolution spectroscopic data, this method takes advantage of advances in AO. It could also be implemented on space-borne spectroscopic observations; however, this requires rotating the spectrograph slit in two perpendicular directions.
The first tests presented here confirm the results obtained in Paper I and show that there is a steep height decrease of the rotational velocity in the low photosphere of the Sun. Due to the large error bars in the measurements, the exact value of this gradient remains uncertain.  The decrease remains significant, reaching $\Delta v=-0.3\pm 0.1$~km~~s$^{-1}$ at $z=75 \pm 16$~km. 
Our measurement accuracy does not allow us to detect any meridional velocity height gradient in the low photosphere below $z = 80$~km, nor to identify significant latitudinal variations of the rotational shear.

It should be noted that a radial gradient in rotational velocity could influence differential rotation measurements \citep[e.g.,][]{roudier2018}, since observations at high latitudes are made at smaller limb distances, i.e., at higher altitudes than measurements at low latitudes. For continuum at 630~nm, the formation height for heliocentric angle $50^\circ$ is about $35$~km. From our measurements, this corresponds to a difference in velocity of about $-0.1$~km s$^{-1}$ , which is about the difference found at that latitude between Doppler measurements and methods based on correlation tracking using continuum images of granulation (see Fig.~1 of \citet{roudier2018}) .  

In a recent work, \citet{2024CHASE} use Dopplergrams obtained by the Chinese H$\alpha$ Solar Explorer (CHASE) to infer the height dependence of the solar differential rotation for photospheric layers above $70$~km up to the chromosphere. From measurements in Si~I (656.06~nm) and Fe~I (656.92~nm) spectral lines, they found an increase of the rotational velocity of about +0.03~km s$^{-1}$ at the equator between 70~km and 250~km. In this work, we found an opposite and steeper gradient of -0.3~km s$^{-1}$ between 0 and 70~km. This would imply an inversion of the radial gradient in a photospheric layer around 70~km.  
These first results must be confirmed by new observations and the accuracy of the radial shear measurement should be improved by analyzing larger statistical samples.
\bigskip

\begin{acknowledgements}
We dedicate this work to the memory of our colleague Didier Laforgue, whose contributions were invaluable. He passed away before the completion of this paper.
This work was supported by the National PNST Program of CNRS / INSU cofunded by CNES and CEA. The THEMIS telescope is operated on the island of Tenerife and yearly funded by the French Centre National de la Recherche Scientifique - Institut National des Sciences de l'Univers: CNRS-INSU. HINODE is a Japanese mission developed and launched by ISAS/JAXA, collaborating with NAOJ as a domestic partner and NASA and STFC (UK) as international partners. The scientific operation of the HINODE mission is conducted by the HINODE science team organized at ISAS/JAXA. This team consists mainly of scientists from institutes in the partner countries. Support for the post-launch operation is provided by JAXA and NAOJ (Japan), STFC (UK), NASA, ESA, and NSC (Norway). 
\end{acknowledgements}

\bibliographystyle{aa}
\bibliography{faurob}

\end{document}